\documentclass[prd,twocolumn,aps,superscriptaddress,nofootinbib]{revtex4-1}
 \usepackage{amsmath}
\usepackage{amssymb}
\usepackage{graphicx}
\usepackage{color}
\usepackage{bm}
\usepackage{times}
\usepackage{hyperref}
\hypersetup{
  colorlinks=true,
  citecolor=blue,
  linkcolor=blue,
  urlcolor=blue}

\usepackage{epsfig}
\usepackage{dcolumn}
\usepackage{float}

\usepackage{epsfig}
\usepackage{dcolumn}
\usepackage[normalem]{ulem}
\def\be{\begin{equation}}
\def\ee{\end{equation}}
\def\bea{\begin{eqnarray}}
\def\eea{\end{eqnarray}}
\def\gsim{\ \rlap{\raise 2pt\hbox{$>$}}{\lower 2pt \hbox{$\sim$}}\ }
\def\lsim{\ \rlap{\raise 2pt\hbox{$<$}}{\lower 2pt \hbox{$\sim$}}\ }
\def\dslash{\kern-4pt \not{\hbox{\kern-2pt $\partial$}}}
\def\pslash{\not{\hbox{\kern-2pt p}}}

\begin{document}

\DeclareGraphicsExtensions{.eps,.ps}

\title{Phenomenological predictions for pentaquark masses from fits to baryon masses}

\author{Pontus Holma}
\email{pholma@kth.se}
\affiliation{Department of Physics, School of Engineering Sciences,
KTH Royal Institute of Technology, AlbaNova University Center, Roslagstullsbacken 21, SE--106 91 Stockholm, Sweden}
\affiliation{The Oskar Klein Centre for Cosmoparticle Physics, AlbaNova University Center, Roslagstullsbacken 21, SE--106 91 Stockholm, Sweden}

\author{Tommy Ohlsson}
\email{tohlsson@kth.se}
\affiliation{Department of Physics, School of Engineering Sciences,
KTH Royal Institute of Technology, AlbaNova University Center, Roslagstullsbacken 21, SE--106 91 Stockholm, Sweden}
\affiliation{The Oskar Klein Centre for Cosmoparticle Physics, AlbaNova University Center, Roslagstullsbacken 21, SE--106 91 Stockholm, Sweden}
\affiliation{University of Iceland, Science Institute, Dunhaga 3, IS--107 Reykjavik, Iceland}

\begin{abstract}
We investigate the mass spectra of exotic hadrons known as pentaquarks. We extend a simple phenomenologi\-cal model based on the G{\"u}rsey--Radicati mass formula for hadrons to include both charmed and bottom baryons as well as to be able to predict masses of pentaquark states including both charm and bottom quark-antiquark pairs. In particular, we perform numerical fits of this model, which includes seven free parameters, to masses of 21 baryons. We find that the model can be well fitted to the experimental values of the baryon masses and observe that the predicted value of about 4400~MeV for the mass of the pentaquark $P_c(4380)^+$ lies within the experimental range reported by the LHCb experiment. In addition, the predicted value of about 4500~MeV for the mass of the pentaquark $P_c(4450)^+$ is close to the experimental value. Finally, in the future, other predicted values for masses of additional pentaquarks could be shown to agree with upcoming experimental results.  
\end{abstract}

\maketitle

\section{Introduction}

The idea of hadrons, which contain more than the minimal quark content ($q\Bar{q}$ or $qqq$), was proposed by Gell-Mann \cite{GellMann:1964nj} in 1964. This includes the possibility of hadrons containing five quarks, which were given the name pentaquark by Gignoux {\it et al.} \cite{Gignoux:1987cn} and Lipkin \cite{Lipkin:1987sk} in 1987. It was not until 2015 when the two first pentaquark states were conclusively observed by the LHCb collaboration in studying $\mathit{\Lambda}_b^0 \rightarrow J/\psi \; K^- \; p$ decays. These two states are denoted $P_c(4380)^+$ and $P_c (4450)^+$. Both of the discovered states have quark content $uudc\bar{c}$ \cite{Aaij:2015tga}. Furthermore, in 2019, additional pentaquark states were discovered \cite{Aaij:2019vzc}. The $P_c (4450)^+$, formerly reported by LHCb, received further confirmation and is observed to consist of two narrow overlapping peaks, named $P_c(4440)^+$ and $P_c(4457)^+$.

Several quantitative models for masses of pentaquarks have been studied in the literature: a selection of these are briefly summarized below. In 2017, a group theoretical classification and prediction for masses of pentaquarks based on the results from LHCb can be found in Ref.~\cite{Santopinto:2016pkp}, where the masses are predicted from a modified mass formula fitted to masses of baryons. In Refs.~\cite{Bijker:2003pm,Bijker:2003fk}, masses and magnetic moments of pentaquarks were obtained in a constituent quark model with a complete classification of $qqqq\bar{q}$ pentaquark states. In Ref.~\cite{Karliner:2003dt}, Karliner and Lipkin proposed a model for pentaquark states to predict their masses, which is based on a model known to reliably deal with both $qq$ and $q\bar{q}$ interactions. In Ref.~\cite{Hosaka:2005ja}, the mass spectrum of pentaquarks based on the quark model and with emphasis on chiral symmetry was presented. In Ref.~\cite{Tazimi:2016hsv}, the authors reduced the complications of studying five-quark systems by considering the pentaquark as being a bound state composed of a baryon and a meson, thereafter solving the Lippmann--Schwinger equation for this system to obtain approximate masses for some pentaquark states. In Ref.~\cite{Ali:2017ebb}, the mass spectrum of $c\bar{c}$ pentaquarks having $J^P = \tfrac{1}{2}^\pm$ for ${\rm SU}(3)_{f}$ multiplets was explored, whose ana\-ly\-sis is based on the two pentaquark states observed in 2015 by LHCb. In Ref.~\cite{Li:2018vhp}, the authors primarily examined pentaquarks of quark content, which are not presently known to exist, and estimated the masses of the said pentaquarks using the chromomagnetic model. Recently, in Ref.~\cite{Ortiz-Pacheco:2018ccl}, the masses of ground-state hidden-charm pentaquarks with $J^P = \tfrac{3}{2}^-$ based on an SU(4) quark model were calculated, which are considerably smaller than the masses of pentaquarks measured by LHCb \cite{Aaij:2015tga,Aaij:2019vzc}.

In the present work, we will primarily consider the simple mass formula for hadrons presented in Ref.~\cite{Santopinto:2016pkp}, which predicts the masses of pentaquarks using numerical fits to masses of baryons. We will also make modifications to this formula in order to predict masses of pentaquarks not examined in Ref.~\cite{Santopinto:2016pkp}. More specifically, we will consider pentaquarks containing $c\bar{c}$ or $b\bar{b}$.

We will study ${\rm SU}(6)_{sf}$ spin-flavor pentaquark configurations, which can be decomposed as ${\rm SU}(3)_f \otimes {\rm SU}(2)_s$. The ${\rm SU}(3)_f$ multiplets are relevant when determining the masses of the pentaquark states. These multiplets will also be part of the baryon data used to perform the numerical fits. We will make use of the Young tableaux technique and let each representation be denoted by $[f_1,\ldots,f_n]_d$, where $f_i$ stands for the number of boxes in the $i$th row of the Young tableau and $d$ is the dimension of the representation. This will follow the classification outlined in Ref.~\cite{Santopinto:2016pkp}, where the main study of the ${\rm SU}(6)_{sf}$ spin-flavor pentaquark configurations was originally performed.

It should be noted that we will focus on the possibility of pentaquarks having $J^P = \tfrac{3}{2}^\pm, \tfrac{5}{2}^\pm$. We do not attempt to describe the binding mechanism for pentaquarks, but in Ref.~\cite{Aaij:2019vzc}, it is clear that $J^P = \tfrac{1}{2}^-$ seems to be the most reasonable option if a pentaquark is indeed a bound system consisting of a baryon and a meson, i.e.~a molecular description. However, as reported in Ref.~\cite{Cho:2019}, the molecular description of pentaquarks is currently the most favored one.

This work is organized as follows. In Sec.~\ref{sec:proc}, we present the simple mass model for hadrons that can predict masses of pentaquarks and describe the numerical fitting procedure used. Then, in Sec.~\ref{sec:nume}, we perform seven numerical fits of this model and state the results of these fits, including predicted masses of pentaquarks. Finally, in Sec.~\ref{sec:summ}, we summarize our main results and conclude based on our presented results.


\section{Model and fitting procedure}
\label{sec:proc}

In this work, we investigate a simple model for hadron mass spectra based on a generalization of the G{\"u}rsey--Radicati mass formula \cite{Gursey:1992dc} performed in Ref.~\cite{Santopinto:2016pkp} for the charm sector, which reads
\begin{align}
    M_H &= \xi M_0 + A S(S+1) + D Y \nonumber\\
    &+ E \left[I(I+1)-\frac{1}{4}Y^2\right] + GC_2 ({\rm SU}(3)_f) + \sum_{i = c,b} F_i N_i \,,
    \label{eq:GR}
\end{align}
where $S$, $Y$, $I$, $C_2({\rm SU}(3)_f)$, $N_c$, and $N_b$ are the spin, hypercharge, isospin, eigenvalue of the ${\rm SU}(3)_f$ Casimir operator, number of charm quarks and antiquarks, and number of bottom quarks and antiquarks of the hadron $H$, respectively. In addition, $M_0$, $A$, $D$, $E$, $G$, $F_c$, and $F_b$ are free para\-meters of the model and the parameter $\xi$ is a scale factor related to the number of particles making up the hadron $H$. It should be noted that Eq.~(\ref{eq:GR}) holds for baryons, but not mesons which follow another but similar mass formula \cite{Gursey:1992dc}. However, we assume Eq.~(\ref{eq:GR}) to hold also for pentaquarks (including charm and bottom quarks) and change $\xi$ accordingly. Thus, the model has either six (including charmed baryons and pentaquarks) or seven free para\-meters (including both charmed and bottom baryons and pentaquarks). For baryons, $\xi = 1$, whereas for pentaquarks, $\xi = 5/3$, since the number of quarks in a pentaquark is five instead of three as in a baryon. The original G{\"u}rsey--Radicati mass formula \cite{Gursey:1992dc} stems from a group theoretical approach and is based on a broken ${\rm SU}(6)$ symmetry. The generalization in Eq.~(\ref{eq:GR}) is a natural extension of the original formula. The strongly broken symmetry of ${\rm SU}(6)$ is modeled and handled by the different free para\-meters in Eq.~(\ref{eq:GR}), especially by $F_c$ and $F_b$. In addition, it should be noted that another generalization of the G{\"u}rsey--Radicati mass formula for the strange sector can be found in Ref.~\cite{Giannini:2005ks}.

Now, we present our fitting procedure. In order to find the best-fit values of the free parameters, we use a $\chi^2$ test, which gives rise to the goodness of fit, by minimizing the corresponding $\chi^2$ function with respect to the free parameters in Eq.~(\ref{eq:GR}). More precisely, for the model with seven free para\-meters, we minimize the following function
\begin{equation}
    \chi^2 = \sum_{H=1}^n \left[ \frac{M_H(M_0, A, D, E, G, F_c, F_b) - M_H^{\rm exp.}}{\sigma_H^{\rm exp.}}\right]^2,
\label{eq:chi2}
\end{equation}
where $n$ is the number of hadrons used, $M_H$ denotes the predicted mass of the hadron $H$ by the generalized G{\"u}rsey--Radicati mass formula (\ref{eq:GR}), and $M_H^{\rm exp.}$ and $\sigma_H^{\rm exp.}$ are the experimental value for the mass of the hadron $H$ and its corresponding experimental error, respectively. For the model with six free parameters, we have to remove the dependence on one of the free parameters $F_c$ and $F_b$, which can be done by enforcing either $F_c = 0$ or $F_b = 0$ in the minimization of Eq.~(\ref{eq:chi2}). Note that Eq.~(\ref{eq:GR}) is linear in all free parameters, and therefore, the minimum of Eq.~(\ref{eq:chi2}) can be found exactly.

To date, there are only three (or four) experimentally known pentaquarks containing $c \bar{c}$ \cite{Cho:2019}, and therefore, there is not enough information to determine the free parameters $M_0$, $A$, $D$, $E$, $G$, $F_c$, and $F_b$, using the experimental data of these pentaquarks. Thus, due to the lack of experimental results on pentaquarks, we fit the free parameters to the values of the known baryon mass spectra and assume that the values of the free parameters will be the same for pentaquarks, i.e.~the parameters are universal. This is an extended (and more rigorous) procedure to the similar approach presented in Ref.~\cite{Santopinto:2016pkp}, which did however not include bottom hadrons. In Tab.~\ref{tab:baryon}, the values of the masses and errors (i.e.~the baryon spectra) that we use in our fits are listed. Since the $\chi^2$ function in Eq.~(\ref{eq:chi2}) weighs the different terms based on their experimental errors, we expect the best-fit values of the free parameters to give an accurate value for the mass of $N(940)$ (i.e.~the neutron), which is the baryon that has the lowest experimental error by far (cf.~Tab.~\ref{tab:baryon}).

\begin{table*}
\centering
\caption{Experimental values and errors for the masses of 21 selected baryons including corresponding quantum numbers for the baryons. All values are taken from Ref.~\cite{Tanabashi:2018oca}. Note that it has been assumed that the ${\rm SU}(3)_f$ multiplet of a bottom baryon is the same as the one for the corresponding charmed baryon, see review ``104.~Charmed Baryons'' in Ref.~\cite{Tanabashi:2018oca}. The 21 selected baryons are all ground-state non-strange baryons, hyperons, charmed baryons, or bottom baryons.}
\begin{tabular}{l c c c c c c c c c c}
\hline\hline
Baryon & Exp.~mass [MeV] & Exp.~error [MeV] & ${\rm SU}(3)_f$ multiplet & $C_2({\rm SU}(3)_f)$ & $S$ & $Y$ & $I$ & $N_c$ & $N_b$  \\ [0.1ex] 
\hline \\[-2.3ex]
$N(940)$ & 939.5654133 & $\pm 5.8 \times10^{-6}$ & $[21]_8$ & 3 & $\frac{1}{2}$ & 1 & $\frac{1}{2}$ & 0 & 0\\ [0.3ex] 
$\Lambda^0$ & 1115.683 & $\pm 0.006$ & $[21]_8$ & 3 & $\frac{1}{2}$ & 0 & 0 & 0 & 0\\ [0.3ex]
$\Sigma^0$ & 1192.642 & $\pm 0.024$ & $[21]_8$ & 3 & $\frac{1}{2}$ & 0 & 1 & 0 & 0\\ [0.3ex]
$\Xi^0$ & 1314.86 & $\pm 0.20$ & $[21]_8$ & 3 & $\frac{1}{2}$ & $-1$ & $\frac{1}{2}$ & 0 & 0\\ [0.3ex]
$\Delta^0(1232)$ & 1232 & $\pm 2$ & $[3]_{10}$ & 6 & $\frac{3}{2}$ & 1 & $\frac{3}{2}$ & 0 & 0\\ [0.3ex]
$\Sigma^{*0}(1385)$ & 1383.7 & $\pm 1.0$ & $[3]_{10}$ & 6 & $\frac{3}{2}$ & 0 & 1 & 0 & 0\\ [0.3ex]
$\Xi^{*0}(1530)$ & 1531.80 & $\pm 0.32$ & $[3]_{10}$ & 6 & $\frac{3}{2}$ & $-1$ & $\frac{3}{2}$ & 0 & 0\\ [0.3ex]
$\Omega^-$ & 1672.45 & $\pm 0.29$ & $[3]_{10}$ & 6 & $\frac{3}{2}$ & $-2$ & 0 & 0 & 0\\ [0.3ex]
\hline \\[-2.3ex]
$\Lambda_c^+$ & 2286.46 & $\pm 0.14$ & $[11]_3$ & $\frac{4}{3}$ & $\frac{1}{2}$ & $\frac{2}{3}$ & 0 & 1 & 0\\ [0.3ex]
$\Sigma_c^0(2455)$ & 2453.75 & $\pm 0.14$ & $[2]_6$ & $\frac{10}{3}$ & $\frac{1}{2}$ & $\frac{2}{3}$ & 1 & 1 & 0\\ [0.3ex]
$\Xi_c^0$ & 2470.87 & ${}_{-0.31}^{+0.28}$ & $[11]_3$ & $\frac{4}{3}$ & $\frac{1}{2}$ & $-\frac{1}{3}$ & $\frac{1}{2}$ & 1 & 0\\ [0.3ex]
$\Xi_c^{\prime 0}$ & 2578.8 & $\pm 0.5$ & $[2]_6$ & $\frac{10}{3}$ & $\frac{1}{2}$ & $-\frac{1}{3}$ & $\frac{1}{2}$ & 1 & 0\\ [0.3ex]
$\Omega_c^0$ & 2695.2 & $\pm 1.7$ & $[2]_6$ & $\frac{10}{3}$ & $\frac{1}{2}$ & $-\frac{4}{3}$ & 0 & 1 & 0\\ [0.3ex]
$\Omega_c^{*0}(2770)$ & 2765.9 & $\pm 2.0$ & $[2]_6$ & $\frac{10}{3}$ & $\frac{3}{2}$ & $-\frac{4}{3}$ & 0 & 1 & 0\\ [0.3ex]
$\Sigma_c^{*0}(2520)$ & 2518.48 & $\pm 0.20$ & $[2]_6$ & $\frac{10}{3}$ & $\frac{3}{2}$ & $\frac{2}{3}$ & 1 & 1 & 0\\ [0.3ex]
$\Xi_c^{*0}(2645)$ & 2646.32 & $\pm 0.31$ & $[2]_6$ & $\frac{10}{3}$ & $\frac{3}{2}$ & $-\frac{1}{3}$ & $\frac{1}{2}$ & 1 & 0\\ [0.3ex]
\hline \\[-2.3ex]
$\Lambda_b^0$ & 5619.60 & $\pm 0.17$ & $[11]_3$ & $\frac{4}{3}$ & $\frac{1}{2}$ & $\frac{4}{3}$ & 0 & 0 & 1 \\ [0.3ex]
$\Xi_b^0$ & 5791.9 & $\pm 0.5$ & $[11]_3$ & $\frac{4}{3}$ & $\frac{1}{2}$ & $\frac{1}{3}$ & $\frac{1}{2}$ & 0 & 1 \\ [0.3ex]
$\Sigma_b^+$ & 5811.3 &  $^{+0.9}_{-0.8}\pm 1.7$ & $[2]_6$ & $\frac{10}{3}$ & $\frac{1}{2}$ & $\frac{4}{3}$ & $1$ & 0 & 1 \\ [0.3ex]
$\Sigma_b^{*+}$ & 5832.1 & $\pm 0.7 ^{+1.7}_{-1.8}$ & $[2]_6$ & $\frac{10}{3}$ & $\frac{3}{2}$ & $\frac{4}{3}$ & 1 & 0 & 1 \\ [0.3ex]
$\Omega_b^-$ & 6046.1 & $\pm 1.7$ & $[2]_6$ & $\frac{10}{3}$ & $\frac{1}{2}$ & $-\frac{2}{3}$ & 0 & 0 & 1 \\ [0.3ex]
\hline\hline
\end{tabular}
\label{tab:baryon}
\end{table*}

\section{Numerical fits and results}
\label{sec:nume}

First, we calculate the free parameters using the same data for the baryon masses as in Ref.~\cite{Santopinto:2016pkp} and compare our results to those presented in that work. Since it is not mentioned in this work the exact method used to obtain the values of these parameters, it is challenging to determine how those values were obtained. Using the values of the free parameters as well as the baryon masses and the corresponding errors presented in Ref.~\cite{Santopinto:2016pkp}, we obtain $\chi^2 \simeq 1.1 \times 10^{15}$, which is a very large value for a $\chi^2$ function. However, if we minimize Eq.~(\ref{eq:chi2}) with respect to $M_0$, $A$, $D$, $E$, $G$, and $F_c$ under the condition $F_b = 0$ and use the data in Ref.~\cite{Santopinto:2016pkp}, we find that $M_0 \simeq 980.8~{\rm MeV}$, $A \simeq 17.08~{\rm MeV}$, $D \simeq -195.1~{\rm MeV}$, $E \simeq 38.08~{\rm MeV}$, $G \simeq 40.68~{\rm MeV}$, and $F \simeq 1378~{\rm MeV}$ with $\chi^2 \simeq 1.220 \times 10^5$, which is about ten orders of magnitude smaller than the value presented in Ref.~\cite{Santopinto:2016pkp}. Therefore, we believe that the values of the six free parameters in Ref.~\cite{Santopinto:2016pkp} cannot constitute the true minimum of the $\chi^2$ function. In addition, our fit leads to the mass of $N(940)$ to be 939.6~MeV, whereas the fit in Ref.~\cite{Santopinto:2016pkp} gives 972.45~MeV, which explains why the $\chi^2$ function is so large for that fit.

Next, we define seven different fits that we perform in this work, which are the following
\begin{itemize}
\item[\bf Fit I:] Data for the 16 baryons in Tab.~\ref{tab:baryon} with $N_c = N_b = 0$ or $N_c = 1$ and 6 free parameters (${\rm d.o.f.} = 16-6 = 10$),
\item[\bf Fit II:] Data for 15 baryons in Tab.~\ref{tab:baryon} with $N_c = N_b = 0$ or $N_c = 1$ [excluding $N(940)$] and 6 free parameters (${\rm d.o.f.} = 15-6 = 9$),
\item[\bf Fit III:] Data for the 16 baryons in Tab.~\ref{tab:baryon} with $N_c = N_b = 0$ or $N_c = 1$, but with errors equal to 1~\% of the experi\-mental mass values, and 6 free parameters (${\rm d.o.f.} = 16-6 = 10$),
\item[\bf Fit IV:] Data for the 13 baryons in Tab.~\ref{tab:baryon} with $N_c = N_b = 0$ or $N_b = 1$ and 6 free parameters (${\rm d.o.f.} = 13-6 = 7$),
\item[\bf Fit V:] Data for the 13 baryons in Tab. I with $N_c=N_b =0$ or $N_b = 1$, but with errors equal to 1 \% of the experimental mass values, and 6 free parameters (${\rm d.o.f.} = 13 -6 = 7$).
\item[\bf Fit VI:] Data for all 21 baryons in Tab.~\ref{tab:baryon} and 7 free parameters (${\rm d.o.f.} = 21-7 = 14$).
\item[\bf Fit VII:] Data for all 21 baryons in Tab. I, but with errors equal to 1 \% of the experimental mass values, and 7 free parameters (${\rm d.o.f.} = 21 -7 = 14$).
\end{itemize}

In Tab.~\ref{tab:fitparam}, we display the results of the minimization of the $\chi^2$ function in Eq.~(\ref{eq:chi2}) for the seven different fits. Basically, the results of Fit~I consist of corrected and updated results compared to the earlier results presented in Ref.~\cite{Santopinto:2016pkp}. We note that Fit~I leads to very similar values of the free parameters as in the case of using the data for the baryon spectra given in Ref.~\cite{Santopinto:2016pkp}, and still $\chi^2/{\rm d.o.f.} \simeq 1.298 \times 10^4$ is a very large number. Despite the fact that the term for $N(940)$ dominates the $\chi^2$ function in Eq.~(\ref{eq:chi2}), the removal of $N(940)$ from the fit does not change the goodness of fit for Fit~II by much compared to Fit~I, since the value of the $\chi^2$ function per degrees of freedom only shrinks to $\chi^2/{\rm d.o.f.} \simeq 1.235 \times 10^4$. In addition, the predicted value for the mass of $N(940)$ becomes 955.6~MeV, which is not very close to the well-measured experimental value. Then, the results of Fit~III show that by increasing the errors from the experimental ones to values being 1~\% of the corresponding experimental masses the value of the $\chi^2$ function per degrees of freedom decreases significantly to $\chi^2 \simeq 2.544$, which actually constitutes a very good fit. However, the fitted value for the mass of $N(940)$ is 956.1~MeV, which is again far away from the experimental value. Next, instead of using mass spectra for charmed baryons, we use mass spectra for bottom baryons. The results of Fit~IV (compared to Fit~I) show that the fitted values of the free parameters are only somewhat different, and the value of the $\chi^2$ function per degrees of freedom increases slightly to $\chi^2/{\rm d.o.f.} \simeq 1.418 \times 10^4$. Again, we use mass spectra for bottom baryons, but now as for Fit~III with errors of 1~\% of the experimental baryon masses. The results of Fit~V lead to $\chi^2/{\rm d.o.f.} \simeq 3.076$, which is much smaller than for Fit~IV (cf.~the situation for Fits~I and III). Thus, we conclude that using errors of 1~\% of the experimental baryon masses in fits to both types of mass spectra (i.e.~mass spectra for charmed or bottom baryons) lead to values of the $\chi^2$ function per degrees of freedom that are considerably smaller than using the experimental errors of the baryon masses. Our conclusion confirms the predicted uncertainty of the generalized G{\"u}rsey--Radicati mass formula already found in Ref.~\cite{Santopinto:2016pkp} for charmed baryons. Following the procedure of Fits~I and IV, we also perform a fit using mass spectra for both charmed and bottom baryons. The results of Fit~VI give $\chi^2/{\rm d.o.f.} \simeq 1.053 \times 10^4$, which is at least smaller than for Fits~I, II, and IV. Finally, the results of Fit~VII based on a fit of seven free parameters to masses of 21 baryons (including both charmed and bottom baryons) render the best fit with a value of the $\chi^2$ function per degrees of freedom given by $\chi^2/{\rm d.o.f.} \simeq 1.908$. For Fit~VII, using three significant digits, the best-fit values of the seven free parameters are $M_0 \simeq 963~{\rm MeV}$, $A \simeq 21.2~{\rm MeV}$, $D \simeq - 175~{\rm MeV}$, $E \simeq 22.4~{\rm MeV}$, $G \simeq 47.1~{\rm MeV}$, $F_c \simeq 1370~{\rm MeV}$, and $F_b \simeq 4820~{\rm MeV}$. In fact, the best-fit values of the six parameters $M_0$, $A$, $D$, $E$, $G$, and $F_c$ are rather similar to the ones of Fit~III, so the effect of including bottom baryons in the fit seems to be mostly encoded in the free parameter $F_b$. For Fits~III, V, and VII (all using errors of 1~\% of the experimental baryon masses), we find low values of $\chi^2/{\rm d.o.f.}$, obviously the lowest being about $1.91$ for Fit~VII.

\begin{table*}
\centering
\caption{Fitted free parameter values for the seven different fits and the corresponding values of the $\chi^2$ function.}
\begin{tabular}{l r r r r r r r}
\hline\hline
Parameter & Fit I & Fit II & Fit III & Fit IV & Fit V & Fit VI & Fit VII \\ [0.1ex] 
\hline \\[-2.3ex]
$M_0$ [MeV] & 980.3 & 971.2 & 965.3 & 1003 & 953.7 & 980.3 & 962.9 \\ [0.3ex]
$A$ [MeV] & 17.08 & 19.50 & 21.78 & 18.29 & 15.82 & 17.15 & 21.17 \\ [0.3ex]
$D$ [MeV] & $-195.1$ & $-179.2$ & $-175.3$ & $-195.2$ & $-177.8$ & $-195.1$ & $-174.7$ \\ [0.3ex]
$E$ [MeV] & 38.07 & 38.04 & $22.54$ & 37.99 & 21.35 & 38.02 & 22.44 \\ [0.3ex]
$G$ [MeV] & 40.85 & 43.32 & 46.14 & 33.11 & 52.07 & 40.82 & 47.06 \\ [0.3ex]
$F_c$ [MeV] & 1376 & 1368 & 1366 & $\times$ & $\times$ & 1376 & 1367 \\ [0.3ex]
$F_b$ [MeV] & $\times$ & $\times$ & $\times$ & 4829 & 4830 & 4842 & 4823 \\ [0.3ex]
\hline \\[-2.3ex]
$\chi^2$ & 129800 &111100 & 25.44 & 99250 & 21.53 & 147400 & 26.71 \\ [0.3ex]
$\chi^2/{\rm d.o.f.}$ & 12980 & 12350 & 2.544 & 14180 & 3.076 & 10530 & 1.908 \\ [0.3ex]
\hline\hline
\end{tabular}
\label{tab:fitparam}
\end{table*}

\begin{figure}
\includegraphics[scale=0.7]{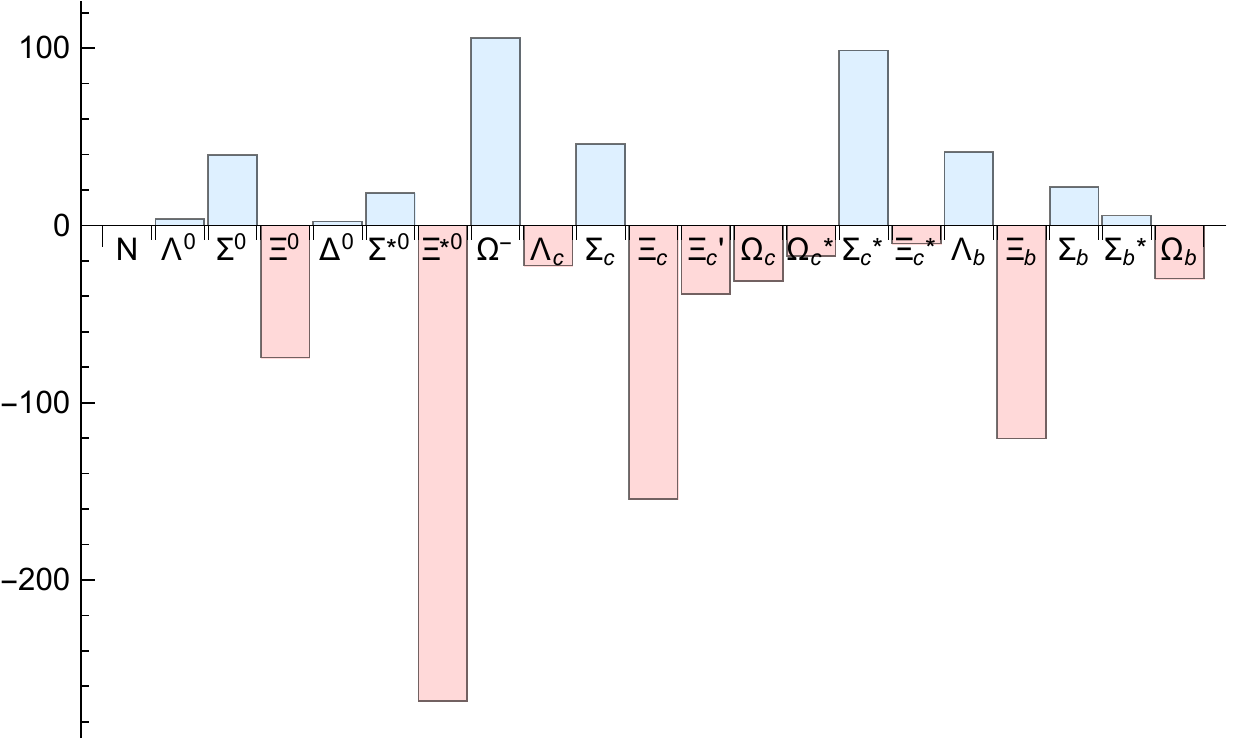}
\caption{Pulls for the 21 fitted baryon masses using the values of the free parameters from Fit~VI.}
\label{fig:pulls}
\end{figure}

\begin{figure}
\includegraphics[scale=0.7]{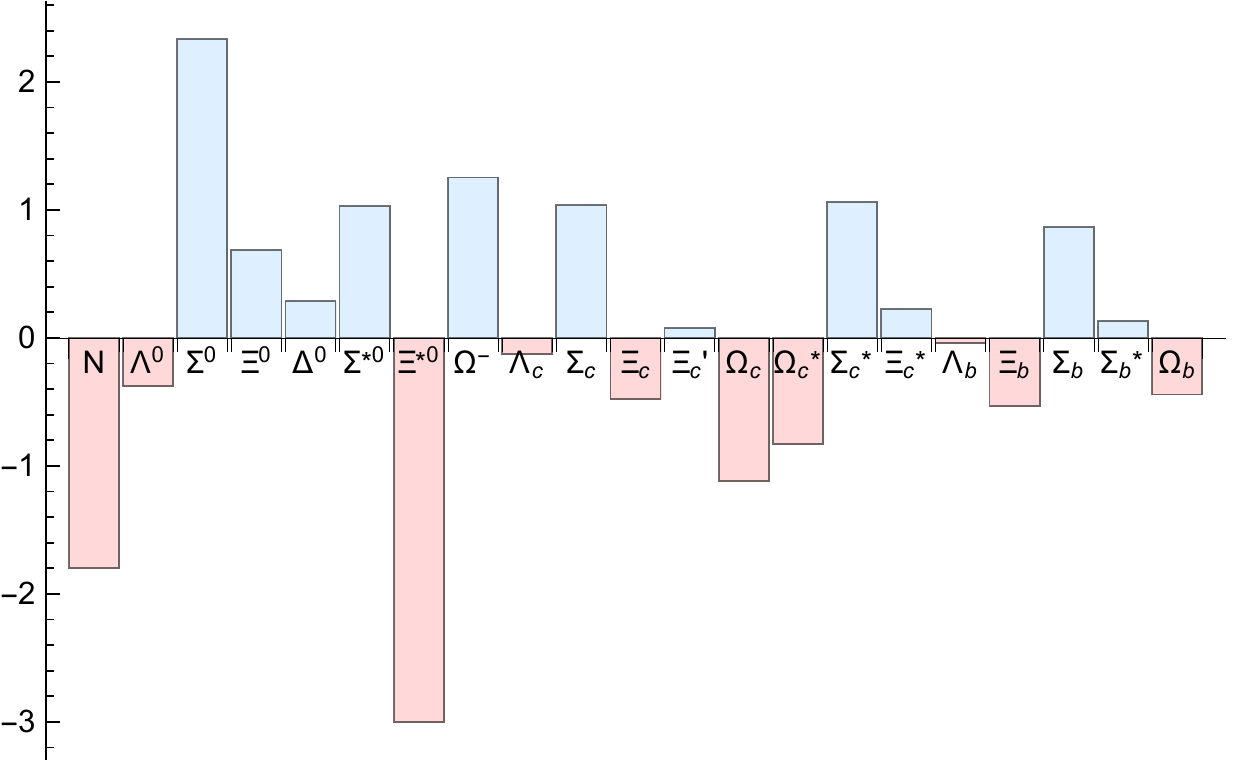}
\caption{Pulls for the 21 fitted baryon masses using the values of the free parameters from Fit~VII.}
\label{fig:pulls1percent}
\end{figure}

In Figs.~\ref{fig:pulls} and \ref{fig:pulls1percent}, we display the pulls for the 21 fitted baryon masses using the values of the free parameters from Fits~VI and VII, respectively. First, in Fig.~\ref{fig:pulls}, we observe that the pull for $N(940)$ is negligible. Furthermore, the two baryons $\Lambda^0$ and $\Delta^0(1232)$ have also small pulls, whereas the largest pull comes from $\Xi^{*0}(1385)$. In general, it is interesting to note that all $\Xi$ and $\Omega_q$ baryons have negative pulls, while all $\Sigma$ baryons and $\Omega^-$ have positive pulls. Second, in Fig.~\ref{fig:pulls1percent}, we observe that the three largest pulls originate from $N(940)$, $\Sigma^0$, and $\Xi^{*0}(1385)$. Due to the choice of the errors being 1~\% of the experimental baryon masses for this fit, the pull for $N(940)$ is naturally non-negligible. Comparing the pulls from the two fits, it is striking to note that the largest pull in both fits comes from $\Xi^{*0}(1385)$. It should be noted that the pulls are generally about 100 times smaller for Fit~VII (see Fig.~\ref{fig:pulls1percent}) than for Fit~VI (see Fig.~\ref{fig:pulls}). Finally, despite the approach based on a simple mass formula, it is encouraging that our and previous similar results estimate to such a good accuracy the baryon mass spectra (the pulls of Fit~VII are at most a couple of per mille of the experimental values for the baryon masses). However, we are aware that this approach is not as sophisticated as more advanced models based on QCD sum rules, various quark models, and effective field theory (see e.g.~Refs.~\cite{Chen:2015moa,Wang:2015epa,Yang:2015bmv,Chen:2016otp,Ali:2016dkf,Park:2017jbn,Ali:2017ebb,Azizi:2017bgs,Yamaguchi:2017zmn,Anwar:2018bpu,Yang:2018oqd,Li:2018vhp,Weng:2019ynv}).

Now, we study five different sets of quantum numbers that could be identified as pentaquark states, which are the following
\begin{itemize}
\item $P^{0+}_q = uudq\Bar{q}$, $P^{00}_q = uddq\Bar{q}$: $S=\tfrac{3}{2}$, $Y = 1$, $I = \tfrac{1}{2}$,
\item $P^{0+}_q = uudq\Bar{q}$, $P^{00}_q = uddq\Bar{q}$: $S = \tfrac{5}{2}$, $Y = 1$, $I = \tfrac{1}{2}$,
\item $P^{1'0}_q = udsq\Bar{q}$: $S = \tfrac{3}{2}$, $Y = 0$, $I = 0$,
\item $P^{1+}_q = uusq\Bar{q}$, $P^{1-}_q = ddsq\Bar{q}$, $P^{10}_q = udsq\Bar{q}$: $S = \tfrac{3}{2}$, $Y = 0$, $I = 1$,
\item $P^{20}_q = ussq\Bar{q}$, $P^{2-}_q = dssq\Bar{q}$: $S = \tfrac{3}{2}$, $Y = -1$, $I = \tfrac{1}{2}$,
\end{itemize}
where $q = c,b$, i.e.~$q\Bar{q}$ is either a charm-anticharm pair or a bottom-antibottom pair. In fact, all above states with $S = 3/2$ belong to the ${\rm SU}(3)_f$ octet, which has $C_2({\rm SU}(3)_f) = 3$ (that is also assumed for the states with $S = 5/2)$. It should, of course, be noted that other sets of quantum numbers for potential pentaquark states could also be studied using Eq.~(\ref{eq:GR}).

In Tab.~\ref{tab:pentaquark}, we present the predicted mass spectrum of pentaquark states containing charm and bottom quark-antiquark pairs for the seven different fits that we have performed, using Eq.~(\ref{eq:GR}) with $\xi = 5/3$ and the best-fit values of the free para\-meters for each of the fits given in Tab.~\ref{tab:fitparam}. Especially, for Fit~VII with three significant digits, the predicted mass spectrum is given by $4400~{\rm MeV}$, $4500~{\rm MeV}$, $4560~{\rm MeV}$, $4600~{\rm MeV}$, and $4740~{\rm MeV}$ for pentaquarks with $c\bar{c}$ and $11300~{\rm MeV}$, $11400~{\rm MeV}$, $11500~{\rm MeV}$, $11500~{\rm MeV}$, and $11700~{\rm MeV}$ for pentaquarks with $b\bar{b}$. In fact, it should be noted that the predicted mass spectra for pentaquarks with $c\bar{c}$ for Fits~III (I) and VII (VI) exactly coincide using three significant digits (see Tab.~\ref{tab:pentaquark}). Thus, we conclude that the fits using errors equal to 1 \% of the experimental baryon mass values do not give much different results than the other fits for the predicted masses of the pentaquarks, which is one of the central topics in this work. Furthermore, our predicted masses of pentaquarks with $c\bar{c}$ (for all seven fits) are all larger than the four ones reported in Tab.~VI of Ref.~\cite{Santopinto:2016pkp}.

\begin{table*}
\centering
\caption{Predicted mass values of pentaquark states for the seven different fits. All values in this table are presented in units of MeV.}
\begin{tabular}{l r r r r r r r}
\hline\hline
Pentaquark states & Fit I & Fit II & Fit III & Fit IV & Fit V & Fit VI & Fit VII \\ [0.1ex] 
\hline \\[-2.3ex]
$P^{0+}_c$, $P^{00}_c$ ($S=\tfrac{3}{2}$) & 4397 & 4397 & 4396 & $\times$ & $\times$ & 4397 & 4395 \\ [0.3ex]
$P^{0+}_c$, $P^{00}_c$ ($S=\tfrac{5}{2}$) & 4482 & 4495 & 4505 & $\times$ & $\times$ & 4483 & 4501 \\ [0.3ex]
$P^{1'0}_c$ & 4573 & 4558 & 4560 & $\times$ & $\times$ & 4573 & 4559 \\ [0.3ex]
$P^{1+}_c$, $P^{1-}_c$, $P^{10}_c$ & 4649 & 4634 & 4605 & $\times$ & $\times$ & 4649 & 4604 \\ [0.3ex]
$P^{20}_c$, $P^{2-}_c$ & 4787 & 4756 & 4747 & $\times$ & $\times$ & 4787 & 4745 \\ [0.3ex]
\hline \\[-2.3ex]
$P^{0+}_b$, $P^{00}_b$ ($S=\tfrac{3}{2}$) & $\times$ & $\times$ & $\times$ & 11320 & 11300 & 11330 & 11310 \\ [0.3ex]
$P^{0+}_b$, $P^{00}_b$ ($S=\tfrac{5}{2}$) & $\times$ & $\times$ & $\times$ & 11410 & 11380 & 11410 & 11410 \\ [0.3ex]
$P^{1'0}_b$ & $\times$ & $\times$ & $\times$ & 11500 & 11470 & 11500 & 11470 \\ [0.3ex]
$P^{1+}_b$, $P^{1-}_b$, $P^{10}_b$ & $\times$ & $\times$ & $\times$ & 11570 & 11510 & 11580 & 11520 \\ [0.3ex]
$P^{20}_b$, $P^{2-}_b$ & $\times$ & $\times$ & $\times$ & 11710 & 11650 & 11720 & 11660 \\ [0.3ex] 
\hline\hline
\end{tabular}
\label{tab:pentaquark}
\end{table*}

It is encouraging to observe that the predicated value of 4400~MeV for the mass of the pentaquark $P_c(4380)^+$ [$uudc\bar{c}$ ($S=3/2$)] lies within the experimental value of $(4380 \pm 8 \pm 29)~{\rm MeV}$ reported by the LHCb experiment \cite{Aaij:2015tga} and listed in the 2018 review of the Particle Data Group \cite{Tanabashi:2018oca}. In addition, the predicated value of 4500~MeV for the mass of the pentaquark $P_c(4450)^+$ [$uudc\bar{c}$ ($S = 5/2$)] is in the close vicinity of the experimental value of $(4449.8 \pm 1.7 \pm 2.5)~{\rm MeV}$.

It should be mentioned that the latest results of the LHCb experiment now confirm that the previously reported pentaquark $P_c(4450)^+$ is resolved into two narrow pentaquark states, i.e.~$P_c(4440)^+$ and $P_c(4457)^+$ \cite{Aaij:2019vzc}. In addition, another narrow pentaquark $P_c(4312)^+$ is discovered, which might or might not be the pentaquark $P_c(4380)^+$, since the analysis is not sensitive to such wide pentaquark states as $P_c(4380)^+$.

Several previous studies have made predictions for pentaquarks containing $b\bar{b}$ and $c\bar{c}$ . A selection of such studies can be found in Refs.~\cite{Chen:2015moa,Azizi:2017bgs,Yamaguchi:2017zmn,Anwar:2018bpu,Yang:2018oqd}. In general, our predicted masses of pentaquarks containing $b\bar{b}$ seem to be larger than the previously reported results.

\section{Summary and Conclusions}
\label{sec:summ}

We have investigated a simple phenomenological model based on an extension of the G{\"u}rsey--Radicati mass formula to predict masses of pentaquarks containing charm-anticharm or bottom-antibottom pairs from numerical fits to masses of charmed and/or bottom baryons, which are reproduced to a good accuracy. We have found that the predicted values for the masses of the pentaquarks $P_c(4380)^+$ and $P_c(4450)^+$ are about 4400~MeV and 4500~MeV, respectively. It should be noted that the latest LHCb experimental results suggest that $P_c(4450)^+$ is resolved into two pentaquarks $P_c(4440)^+$ and $P_c(4457)^+$. Furthermore, our predicted value of 4500~MeV for $P_c(4450)^+$ (or $P_c(4440)^+$ and $P_c(4457)^+$) assumed that the eigenvalue of the ${\rm SU(3)}_f$ Casimir operator is equal to 3 (as for $P_c(4380)^+$, which belongs to the ${\rm SU(3)}_f$ octet). However, this assumption might be uncertain.

Our model could also be used to predict masses for other potential pentaquark states, given their quantum numbers, for which no experimental data exist today. In addition, using $\xi = 4/3$ in our model, we could predict masses for hadrons consisting of four quarks, i.e.~so-called tetraquarks.

\begin{acknowledgments}
T.O.~acknowledges support by the Swedish Research Council (Vetenskapsr{\aa}det) through Contract No.~2017-03934 and the KTH Royal Institute of Technology for a sabbatical period at the University of Iceland.
\end{acknowledgments}

\vfill

\end{document}